\documentclass{elsart}

\usepackage{epsfig}

\usepackage{amssymb}

\begin{document}

\begin{frontmatter}


\thanks[label2]{Corresponding author}

\title{A 15$^\circ$ Wide Field of View Imaging Air Cherenkov Telescope}


\author[Munich]{R. Mirzoyan\thanksref{label2}},
\author[Copenhagen,Potsdam]{M.I. Andersen\thanksref{label2}},
\address[Munich]{Max-Planck-Institut f\"ur Physik, F\"ohringer Ring 6, 
D-80805 M\"unchen, Germany}
\address[Copenhagen]{Dark Cosmology Centre, Niels Bohr Institute,
        University of Copenhagen, \\
        ~Juliane Mariesvej 30, DK-2100 Kbh \O, Denmark}
\address[Potsdam]{Astrophysikalisches Institut Potsdam,
        An der Sternwarte 16, D-14482 Potsdam, Germany}
%
%


\maketitle

\begin{abstract}
Contemporary imaging air Cherenkov telescopes (IACT)
for ground-based very high energy (VHE) gamma-ray astronomy
have prime focus optical design. Typically these telescopes
have a $(2-4)^{\circ}$ wide field of view (FoV).
They use F/0.7-F/1.2 optics and provide $(3-10)^{'}$ resolution in the FoV.
Generally, a well designed telescope that includes more than one optical
element will offer some advantages not available in prime
focus designs, such as a wider FoV, a more compact size, a higher and 
more homogeneous resolution
and a lower degree of isochronous distortion
of light rays focused onto the focal plane.
Also, they allow monitoring the gamma-ray activity in a
sizeable portion of the sky in a single observation.
This would allow one to perform a sensitive all-sky survey in a
relative short time.
We present an F/0.8 $15^\circ$ wide FoV telescope design, which provides
a high and near uniform resolution and low isochronous distortion
across the entire FoV.

\end{abstract}

\end{frontmatter}

\section{Introduction}

The technique of employing Imaging Air Cherenkov Telescopes (IACTs) has been 
successfully used for ground based gamma-ray astronomy for about
two decades, revealing more than 70 sources in the course of this 
period. The successful race started with the discovery of the first
TeV gamma-ray signal from the Crab Nebula - now regarded as the standard 
candle in gamma-ray astronomy - by the Whipple Collaboration in 1989
\cite{whipple}. Over 20 years following that event, the sensitivity 
of the IACTs has improved dramatically, leading to a large increase 
in the rate of scientific discovery made with these instruments.
In the beginning a few tens of hours were needed to detect 
a significant signal from the Crab Nebula, whereas today{'}s 
installations operating in the same energy range require only 
a few minutes for the same signal strength.
The main 
improvements are essentially due to a) a finer pixel size of photo sensors 
in the imaging camera, b) an improved trigger, c) a larger size of telescopes
and improved optics 
providing stronger signals and revealing more structures in the images 
(helps to further suppress the backgrounds) and 
d) the use of multiple telescopes operating in 
coincidence mode (the so-called stereo mode of observations).
However, the field of view (FoV) of IACTs has not undergone a similar evolution.
The largest FoV telescope had a field of about $7^\circ$ 
\cite{sinitsyna}.
Contemporaneous IACTs typically cover a $(2-4)^{\circ}$ wide FoV.
A wider FoV would enable sensitive
all-sky surveys to be conducted within a relatively short time frame.
In optical astronomy the trend to larger FoV is evident, just to mention 
the LSST \cite{lsst} and the Pan-starrs projects \cite{panstarrs}.
An interesting example of the use of even 
moderately wide FoV telescopes has been demonstrated by the H.E.S.S.
collaboration. While performing observations of scheduled astronomical targets, 
H.E.S.S. has discovered several 
new sources in the $\sim 4^\circ$ effective FoV of their instrument.
Subsequently, when scanning the galactic plane, multiple sources were
found in the $\sim 300$ square degree band surveyed by
H.E.S.S. - i.e. the region scanned was much larger than a 
single H.E.S.S. FoV.
\cite{hess}.

Along with the advantages of the wide FoV there are 
also a number of drawbacks, such as: i) compared to currently 
used simple prime focus constructions, 
they have a more complex optical and mechanical 
design, ii) the imaging camera will have a large transverse size and 
thus can vignet a significant fraction of the mirror and iii) the 
imaging camera will be composed of a very large number of light 
sensors and one will therefore need a large number of readout 
channels. These factors tend to make a wide FoV telescope expensive.
In the following we shall present a concept for a wide FoV IACT, for which
the complications due to i) and ii) are minimal. The challenge of building
a camera with a large number of channels cannot be by-passed.
\section{Wide FoV}
For the successful operation 
of an IACT, one needs to provide a relatively high optical 
resolution to efficiently select the rare gamma shower images from the few 
orders of magnitude more frequent images induced by hadron 
(background) showers. Although the images of gamma showers 
tend to be small in size compared to those of hadrons, 
such a selection is not straightforward
because the distributions of the parameters used to describe
their images (see, for example, \cite{carmona}) overlap significantly.
The differences in shape parameters of gamma 
and hadron images are in the range of $(0.1-0.2)^{\circ}$ for the TeV 
energy range and they are a few times less for the (sub) 100~GeV energy 
range. Therefore, for a successful image discrimination
the telescope shall provide a Point Spread Function 
(PSF) that is $\leq (0.1-0.2)^{\circ}$ for the TeV energy range
and a few times less for the (sub) 100~GeV range.
The simplest and most straightforward way to design a large FoV 
telescope is using the prime-focus design, i.e. with 
just a single (primary) mirror surface of a 
required minimum $f/D$.
Five telescopes of different prime-focus designs were simulated in
\cite{schliesser_mirzoyan}. 
In that study ten optical resolutions in the range of
$(0.01-0.1)^{\circ}$ RMS, with a step size of $(0.01)^{\circ}$, 
were simulated.

It has been shown in \cite{schliesser_mirzoyan}, for example, that 
by using a F/2.7 optics, 
one can design a $10^{\circ}$ wide FoV telescope of 
parabolic design that can provide a resolution of $0.05^\circ$
everywhere in the FoV. 
In the same study it was shown that a Davies-Cotton telescope of 
F/2.5 and even a F/2 optics of elliptical design can provide the 
same $10^{\circ}$ wide FoV at a resolution of $0.05^\circ$, 
albeit at the expense of a higher degree of 
isochronous distortion.
In the recent work \cite{vassiliev} the authors described an
interesting 
$15^{\circ}$ wide FoV aplanatic two mirror telescope design
for gamma-ray astronomy. 
They also showed that one
can improve the angular resolution for gamma events
as the optical resolution in the FoV approaches a limit
of $1^{'}$. 
An alternative way of constructing a wide angle optics is to follow 
the design of the EUSO mission \cite{petrolini}, 
which has refracting optics that
allows a full FoV of $60^{\circ}$ 
or even larger. The GAW telescope for TeV gamma astronomy is following 
that design in their construction \cite{GAW}. Two double-sided Fresnel 
lenses were planned to be used in the optical design of EUSO. 
The disadvantages of that design were considered to be the relatively high 
light losses, especially for relatively large incident angles of light. 
Also, distortion of images because 
of scattering of light by the Fresnel lenses must be carefully taken 
into account. One needs to construct 
the refractive optics from materials that for the given thickness do not 
substantially absorb the short-wave near UV light in the wavelength 
range of 330-400 nm.
A variation of the EUSO-type solution could be to construct a stationary
wide FoV telescope or a telescope that includes some kind of 
secondary optics.

\subsection{Vignetting in prime focus design}
The plate scale of a telescope, giving 
the ratio of angular distance on the sky to length
in the focal plane (in deg./m), is

\begin{equation}
\delta = \frac{57.3}{f}
\end{equation}

where $f$ is the telescope focal length.
For a given detector field of view $\theta$, the diameter
of the detector assembly is

\begin{equation}
d = \frac{\theta}{\delta}
\end{equation}

For a prime focus telescope of a given focal ratio, $F = f/D$,
where D is the telescope diameter, vignetting factor,
i.e., shadowing, of the detector assembly is given by

\begin{equation}
Vignetting = \left( d / D \right)^2 =
\left( \frac{\theta \times f}{57.3} / D \right)^2 =
\left( \theta / 57.3 \right)^{2} \times F^{2}
\end{equation}

Thus, for example, a telescope with $F/2$ and $\theta=15^\circ$
will have a vignetting factor of 27~\%, whereas a camera with a field of view
of $\theta=10^\circ$ will have a vignetting factor of just 12~\%.
We have mentioned above that prime focus telescopes of F/2-F/2.7 of 
a few different designs can provide a FoV of $10^\circ$
with the desired optical resolution. For simplicity 
let us consider an F/2 design. In the case of F/2 design, when compared 
with the F/1 case, we see that:
 
\begin{itemize}

\item the imaging camera will be 2 times further away from the mirror,

\item the imaging camera pixels must have a 2 times larger linear size,

\item the imaging camera weight may increase by more than a factor of 4,

\item the camera support mechanics must be significantly stiffer and

\item the vignetting by the camera will increase by a factor of $\approx$ 4.

\end{itemize}
 
There are therefore several reasons to consider designs which are faster.
Generally, a well designed telescope that includes more than one optical 
element will offer some advantages not available in the case of prime 
focus designs. Those advantages could be a) the wider FoV, b) a higher 
and more homogeneous spatial resolution across the FoV, c) faster 
optics/more compact size and d) lower isochronous distortion.
In the following we want to concentrate on a specific wide FoV telescope 
solution that comprises more than one optical element.
\section{The Schmidt Telescope}

Of all telescope designs, the Schmidt telescope, and solutions 
derived from it, provides the widest FoV.
Specifically, Schmidt type systems provide by far the largest number
of focal plane spatially resolved pixels on the focal plane 
per optical element \cite{schroeder}.
Moreover, it is possible to work at very fast F-ratios, below F/1, 
thereby minimizing the obscuration and weight of the camera, and the 
overall size and weight of the system. 
We have developed
a design of a 7~m Schmidt telescope that is suitable as an IACT.
Our design consists of simple optical components, is
compact (low $f/D$) and is realistic to implement.
The primary design characteristics of the telescope
are given in table \ref{tel_description.tab}. The optical parameters
of the telescope are listed in table \ref{Schmidt_parameters.tab}.

The classical Schmidt telescope uses a spherical mirror and 
an aspheric refractive corrector plate, normally referred to as the
'Schmidt corrector', which is located at the centre of curvature of 
the mirror. The Schmidt telescope has a curved focal plane, which is
con-focal with 
the mirror. The entrance pupil (the stop) is located at the Schmidt corrector. 
In order to accept light without vignetting from directions that are 
relatively far away from the optical axis, the mirror must be 
somewhat larger than the Schmidt corrector.
Thus, for a given incidence angle, only a part of the mirror, 
equivalent to the the size of the Schmidt corrector, is used.

The Schmidt corrector pre-deforms the impinging 
wavefront so that after reflection on the spherical mirror,
it is free of spherical aberration. As the only 'on-axis' aberration of
a spherical mirror is spherical aberration, the Schmidt telescope
is therefore nominally aberration free at the wavelength for which the
the Schmidt corrector is optimized. At an off-axis field position,
the Schmidt corrector makes some angle with the chief ray, 
the chief ray being defined as the ray that passes through the centre of
the entrance pupil, i.e. the centre of the Schmidt corrector.
Thus, while the spherical mirror appears the same for any field position,
the Schmidt corrector will be at an angle to the beam for an off-axis field
position. It will therefore not correct perfectly for spherical aberration
at off-axis field positions. However, because the Schmidt corrector
is a thin plate, aberrations increase very slowly with the field angle.
We see that the Schmidt telescope is free of 3rd order coma and astigmatism,
exactly because the entrance pupil/corrector is placed
at the centre of curvature of the spherical mirror.

A simplified version of the Schmidt-type telescope is used by the 
AUGER collaboration for their air fluorescence telescopes: the 
corrector plate is replaced by an aperture diaphragm, combined with
an annular Schmidt corrector \cite{klages}.
This aperture eliminates the 3rd order coma. The remaining spherical 
aberration is acceptable for the given $f/D$, and satisfactory 
for the requirements of fluorescence telescopes.

\section{The layout of a IACT Schmidt telescope}
In a Schmidt telescope, 
the corrector is a very weak aspheric transparent optical element.
Because the corrector is weak, chromatic effects are moderate
for a telescope with $f/D$ $\approx$ 1.

\begin{figure}[t]
\begin{center}
\includegraphics[totalheight=10cm]{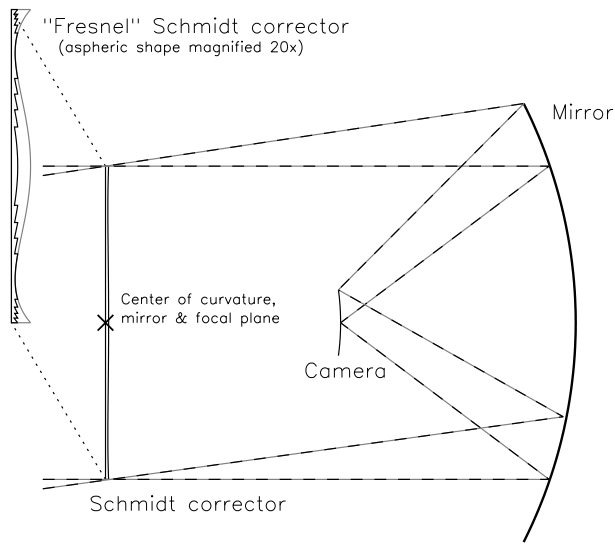}
\end{center}
\caption{\it 
Layout of the Schmidt-type IACT. The mirror and the focal plane have
their centre of curvature at the centre of the corrector plate.
In the insert in the upper left corner, the Schmidt corrector is shown
with the aspheric shape magnified by a factor of 20. Both the nominal
corrector (shaded line) and a Fresnel version is shown.
\label{Mirzoyan_Andersen_fig1} }
\end{figure}
We have developed a specific Schmidt type design, optimizing it 
for the use as wide FoV IACT (Fig. \ref{Mirzoyan_Andersen_fig1}),
using the ZEMAX optics design software.
The design has an F-ratio of 0.80, an entrance aperture 
of 7~m, a total length of 11.2~m and a FoV of $15^\circ$ diameter, with 
a polychromatic image quality that is well below $1^{'}$ RMS radius 
across the entire field. This is achieved with a corrector of acrylic 
plastic and a weakly aspheric mirror.

\begin{table}[h]
\caption{IACT Schmidt telescope primary design parameters.}
\vspace{4mm}
\begin{center}
\begin{tabular}{l l} \hline
Diameter          & ~~~7.0~m                  \\
F-ratio           & ~~~0.8                    \\
Focal length      & ~~~5.6~m                  \\
Field of View     & ~15$^\circ$               \\
Resolution (RMS)  & $<$~1{\mbox{$^{\prime}$}} \\
An-isochronicity  & $\leq$ 0.03~ns            \\
\hline
\end{tabular}
\vspace{2mm}
\label{tel_description.tab}
\end{center}
\end{table}

\begin{table}[h]
\begin{center}
\caption[]{Optical parameters of a 7~m F/0.8 15$^\circ$ FoV Schmidt telescope.}
\vspace{4mm}
\begin{tabular}{l c c c c c} \hline
Component   & Radius of  & Axial  &Diameter& 4th order & 6th order
\vspace{-2mm} \\
          & curvature  & spacing&        & aspheric & aspheric
\vspace{-2mm}\\
          & $[$m$]$   &$[$m$]$   &$[$m$]$&$[$m$^{-3}]$&$[$m$^{-5}]$\\
\hline
Corrector \vspace{-2mm} \\
~~surface 1 &$\infty$    &    \multicolumn{1}{r}{0.0250}~  & 7.00   &\vspace{-3mm}  \\  
~~surface 2 & \multicolumn{1}{r}{-103.00}~   & \multicolumn{1}{r}{10.7600}~  &  7.00   & 3.20$\times 10^{-4}$&  ~~6.75$\times 10^{-6}$ \vspace{2mm} \\
Mirror    &  \multicolumn{1}{r}{-10.88}~   & \multicolumn{1}{r}{-5.2885}~  & 9.95    & 8.30$\times 10^{-7}$& -8.50$\times 10^{-8}$\\
Camera    &   \multicolumn{1}{r}{-5.60}~   &          & 1.47 &       &       \\ 
\hline
\end{tabular}
\vspace{-4mm}
\label{Schmidt_parameters.tab}
\end{center}
\vspace{4mm}
\end{table}

\begin{figure}[h!]
\begin{center}
\includegraphics[totalheight=7cm]{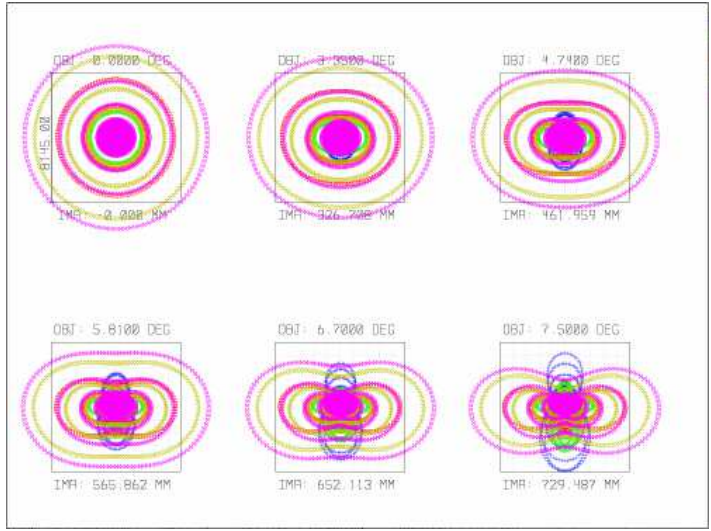}
\end{center}
\caption{\it 
Spot diagram for 6 field positions, from on-axis to $7.5^\circ$.
The box size is 5 arcmin. \label{Mirzoyan_Andersen_fig2} }
\end{figure}

\begin{figure}[h!]
\begin{center}
\includegraphics[totalheight=7cm]{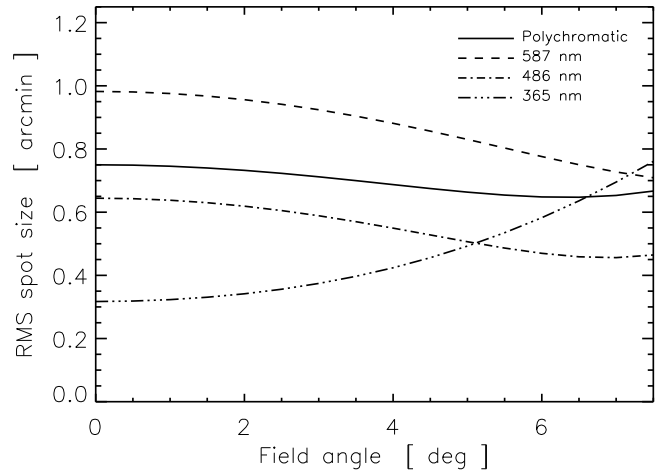}
\end{center}
\caption{\it 
RMS spot size vs. FoV angle. The solid line is for polychromatic light.
\label{Mirzoyan_Andersen_fig3} }
\end{figure}

The optical parameters corresponding to the design are given in
table \ref{Schmidt_parameters.tab}.
Each row of this table fully specifies an optical surface in the telescope.
The first column identifies the component. The refractive corrector naturally has two
surfaces, while the mirror and focal plane are single surfaces.
The second column specifies the radius of curvature of the surface,
with a negative sign signifying that the center of curvature is located 
towards the object.
The third column gives the distance between the vertexes of the current
surface and the next surface, i.e., the spacing of surfaces along the
optical axis.
The 4th column gives the diameter (clear aperture) of the surface.
The 5th and 6th column gives the 4th and 6th order polynomial
coefficients of the corresponding aspheric surfaces.
The aspheric deformation of the mirror is very weak and can be obtained
by optimally tilting segments of a segmented
spherical mirror, as long as the segment size is not more than about 60~cm.
The nominal isochronous distortion is less than 0.01 ns anywhere in the field.
Spot diagrams are shown in Fig. \ref{Mirzoyan_Andersen_fig2} and the
corresponding RMS spot sizes vs FoV angle are given in Fig.
\ref{Mirzoyan_Andersen_fig3}. We note that good imaging characteristics
are naturally linked to low isochronous distortion through Fermat's principle.

\begin{figure}[h!]
\begin{center}
\includegraphics[totalheight=7cm]{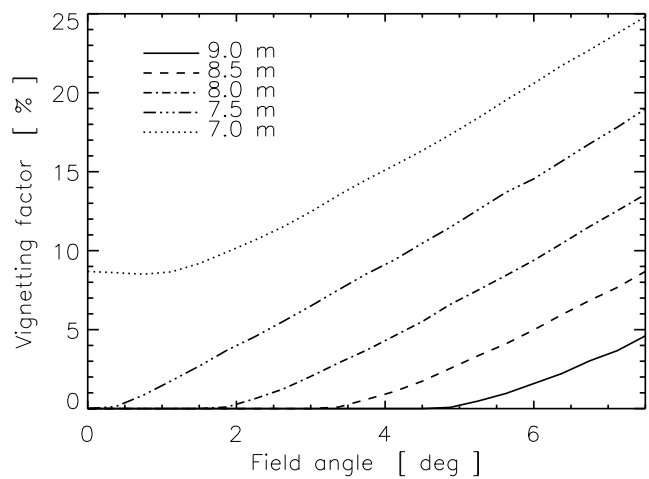}
\end{center}
\caption{\it Vignetting as a function of mirror size.
\label{Mirzoyan_Andersen_fig4} }
\end{figure}

The physical length of a Schmidt telescope is twice its focal length, 
in our design equivalent to a F/1.6 prime focus telescope, which is 
not, by a large margin, capable of delivering a comparable FoV 
with a similar image quality and isochronicity. Because of the very 
fast F-ratio, the camera has a diameter of less than 1.5~m, despite the 
large FoV. The resulting vignetting is less than 5~\%. 
If the entire $15^\circ$ FoV is to be fully illuminated
by the light passing through the Schmidt corrector, the mirror 
must have a diameter of 9.95~m, implying that only 50\% of the mirror 
surface is "actively used" to observe a given point in the sky. By allowing
for some vignetting, the mirror diameter can be reduced down to 7~m.
This is illustrated in Fig. \ref{Mirzoyan_Andersen_fig4}. 
A useful compromise could be a mirror with a diameter of 8~m, 
which would result in 13-14\% vignetting at the very edge of the field.
\section{The practical implementation of a Schmidt IACT}
The mirror of our baseline design is 
essentially the same as those
already implemented
in large Cherenkov telescopes, such as for example MAGIC,
H.E.S.S. or VERITAS, except that
the specification of the alignment of 
the mirror segments is a factor of two to three tighter, 
in order to match the nominal performance of the design. Our Schmidt 
design is ideally suited for implementing an auto-collimation system 
for closed loop control of the mirror alignment. If a light source 
is located at the centre of curvature of the mirror, which is by 
design also the vertex of the corrector plate, light reflected from 
the mirror should return to this point. Any deviation from this 
signifies a deviation from the nominal shape of the mirror. It is 
straightforward to construct a device with a single light source and a 
single camera, which will be able to monitor all mirror segments 
that are not obscured by the camera (about 88\% of the segments), 
in real time, ensuring that the high resolution of the telescope 
can be maintained under all conditions.

\begin{figure}[h!]
\begin{center}
\includegraphics[totalheight=7cm]{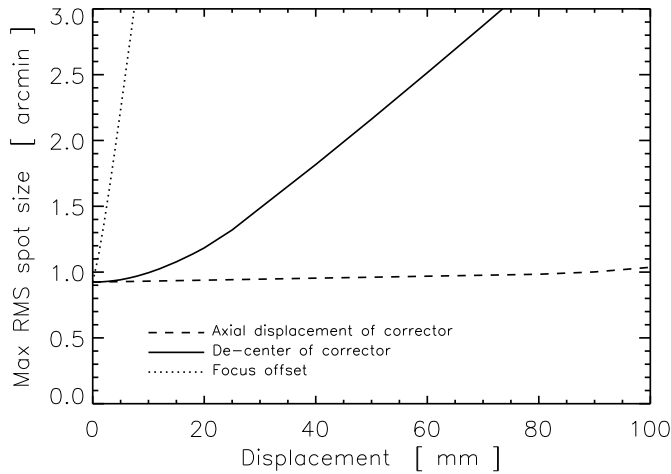}
\end{center}
\caption{\it The maximum RMS spot size over the field of view, as a function of
corrector radial displacement, longitudal displacement and focus offset.
\label{Mirzoyan_Andersen_fig5} }
\end{figure}

The Schmidt corrector is very forgiving with respect to misalignment. 
The centre of the corrector plate should nominally be located at 
the centre of curvature of the mirror, and it should be perpendicular 
to the optical axis. The given design allows for a shift along the optical 
axis of 90~mm and a decenter of 10~mm,
without increasing the spot size beyond $1^{'}$ RMS anywhere in the field. 
This is illustrated in Fig. \ref{Mirzoyan_Andersen_fig5}, where the
consequence of a focus offset is also shown. Tilts of the corrector by
more than one degree are required, before the spot size increases to beyond
$1^{'}$ RMS.
The Schmidt corrector does therefore not pose any new stringent demands on
alignment. The main challenge, with respect to alignment, remains the correct
focussing of the camera across the wide FoV.

The corrector plate is an element which has 
as yet
not been implemented 
on a scale comparable to what is required for the optical system discussed 
here. The corrector plate has a maximum thickness of 17~mm, which would 
imply significant attenuation of the UV radiation. The most practical 
solution is to implement the corrector as a Fresnel like 
lens, whereby 
the thickness of the acrylic corrector can be minimized. Specifically, we consider 
to bond acrylic wedge segments onto the downwards facing surface of a
substrate of 5~mm thick Borofloat sheets.
This would allow for good UV transparency, even below 330~nm.
Both acrylic plastic and Borofloat are inexpensive materials 
which are produced on an industrial scale in large dimensions. 
We note that the use of a Fresnel-like lens implies 
increased isochronous distortion, to a level of 0.03 ns, i.e., a level
which is still very acceptable. Because the aspheric corrector lens is 
very weak, implementing it as a Fresnel lens implies vignetting and scattering
of light on a level well below 0.1\%, even at the edge of the field. 
Thus, the use of a Fresnel lens for the corrector plate does not have
disadvantages that affect the performance of the telescope on a measurable
scale. 
The large Fresnel Schmidt corrector should be implemented as a segmented lens,
where
the segments of a size of $\sim 50~cm$ can be held in a 
spider's web like mesh made from a light-weight material. This 
mesh will further introduce some vignetting, at the level of a few percent
Because the position of the Schmidt corrector is very forgiving, there
is no need for an active control of the segmented corrector.
The dominant aberration in the telescope is chromatic aberration in the
corrector. For the field size and F-ratio used here, this has no practical
importance. It is possible to increase the field of view, and/or use a faster
design, by introducing an achromatic corrector plate. This could be done
by using a combination of Polystyrene and Acrylic plastic materials.
The main challenge would lie in maintaining a high UV transmittance.
With an achromatic corrector plate, a FoV of up to 25 degrees would become
feasible.
\section{A short discussion on the cost of a Wide FoV telescope}
The telescope suggested above has $1^{'}$ resolution anywhere
in the $15^\circ$ wide FoV. The physical size of $1^{'}$ corresponds
to $\sim 1.6~mm$ in the focal plane. This hints
at the possible size of the light sensors that can be used in such
a high resolution system. The currently operating or planned IACTs 
use $\sim 10^{3}$ pixels in their imaging cameras of a few
degree aperture. The $28~m$ diam. H.E.S.S.II telescope will have 
2048 pixels in its $3.2^{o}$ FoV
camera \cite{HESS-II}. 
The wide FoV telescope may need, 
depending on the pixel size and selected FoV, about two orders
of magnitude more pixels for the camera. If the light sensor element 
(that may include a light collector) size is about $5~mm$ one will 
need $7\times 10^{4}$ pixels, and if the
light sensor element size is $10~mm$ one will need 
$\sim 1.8 \times 10^{4}$ pixels for covering a $15^\circ$ wide FoV. 
Usually the readout and the camera of a telescope are
the most expensive items in the total cost. Therefore it will
be mandatory to look for a cost-efficient light sensors and 
readout system. Multi-channel bialkali PMTs or
UV enhanced SiPMs can be used as light sensors. 
The relatively high cost of a wide FoV telescope can be seen as 
compensated by the fact that one will need only one mechanical mount
for covering a huge area in the sky.   
\section{Summary and Conclusions}
We have presented the principal design of a 7~m wide FoV IACT,
which has excellent
imaging characteristics over a $15^\circ$ field diameter.
The basic design is that of a Schmidt type telescope, F/0.8,
i.e., comparatively fast.
This design allows one to obtain an optical 
spot size of $1^{'}$ RMS everywhere in the field, with an isochronous distortion
below 0.03~ns  in case a Fresnel lens is used as a corrector.
It is straightforward to scale this design to larger apertures. The only
aspect which changes by scaling is the isochronous distortion. For a 20~m
diameter telescope, the isochronous distortion amounts to 0.06~ns.
The limiting factor in the baseline design proposed here is chromatic
aberration in the corrector plate. This can be overcome by implementing
an achromatic corrector plate. The main challenge will however lie in
filling the focal plane with detectors which would fully utilize the
resolution provided by the telescope.

\section{Acknowledgements}
We are grateful to Ms. S. Rodriguez for critically reading this 
manuscript. We thank the anonymous referee for providing comments
which have greatly helped to clarify the paper.

\end{document}